# Chat Control or Child Protection?

Ross Anderson
Universities of Cambridge and Edinburgh
Foundation for Information Policy Research

Ian Levy and Crispin Robinson's position paper "Thoughts on child safety on commodity platforms"[1] is to be welcomed for extending the scope of the debate about the extent to which child safety concerns justify legal limits to online privacy. Their paper's context is the laws proposed in both the UK and the EU to give the authorities the power to undermine end-to-end cryptography in online communications services, with a justification of preventing and detecting of child abuse and terrorist recruitment. Both jurisdictions plan to make it easier to get service firms to take down a range of illegal material from their servers; but they also propose to mandate client-side scanning – not just for known illegal images, but for text messages indicative of sexual grooming or terrorist recruitment. In this initial response, I raise technical issues about the capabilities of the technologies the authorities propose to mandate, and a deeper strategic issue: that we should view the child safety debate from the perspective of children at risk of violence, rather than from that of the security and intelligence agencies and the firms that sell surveillance software. The debate on terrorism similarly needs to be grounded in the context in which young people are radicalised. Both political violence and violence against children tend to be politicised and as a result are often poorly policed. Effective policing, particularly of crimes embedded in wicked social problems, must be locally led and involve multiple stakeholders; the idea of using 'artificial intelligence' to replace police officers, social workers and teachers is just the sort of magical thinking that leads to bad policy. The debate must also be conducted within the boundary conditions set by human rights and privacy law, and to be pragmatic must also consider reasonable police priorities.

**Introduction**

Governments have long tried to limit civilian use of cryptography to protect their surveillance capabilities. From export controls during the Cold War, policy moved to mandating exceptional access to keys during 1990s ('Crypto War I') through supply-chain sabotage in the 2000s[2] to law-enforcement demands in 2015 that Apple break iPhone cryptography. The ostensible justification has swung from child protection in the 1990s to terrorism after 9/11 and back to child protection. Since about 2018 the main narrative of law-enforcement and intelligence agencies has been that the end-to-end cryptography in messenger products such as WhatsApp makes life too easy for sexual predators[3], while the introduction of end-to-end encryption in Facebook Messenger is now claimed to pose an additional risk to children. Draft laws have now been introduced in the UK and EU parliaments to enable government agencies to mandate surveillance technology in the name of child protection, though in both cases terrorism has also been advanced as a justification.

---

[1] Arxiv 2207.09506, July 2022
[2] Operations Bullrun and Edgehill, disclosed by Ed Snowden, along with illegal bulk interception
[3] Ian Levy and Crispin Robinson, "Principles for a More Informed Exceptional Access Debate", Lawfare, Nov 29, 2018

There has been significant pushback from industry, and European civil society groups are campaigning vigorously against "Chatcontrol", as the proposal has become known in Brussels. There is wide concern that, just as the move to centralised services such as Gmail and Facebook led the agencies to create a capability to search these services, so also the move to edge computing will lead to edge surveillance, and end-to-end encryption will lead to a government search engine in all our devices. A leaked EU document revealed in 2020 that the search for illegal images of children would be the first argument[4], although the internal discussion touched on terrorism too. The UK Online Safety Bill also mentions terrorism, while the EU's Child Sex Abuse Regulation leaves that for later. Both extend the scope of scanning from illegal images to text messages, and from server-based communications to encrypted messaging in the case of child safety. Yet the European courts prohibit pervasive surveillance without warrant or suspicion. The relevant judgments come not just from the Luxembourg court (whose jurisdiction the UK left after Brexit) but from the Strasbourg court (whose jurisdiction still extends to the UK).

**The Levy-Robinson taxonomy**

Ian Levy is Technical Director of NCSC, a GCHQ division offering protective security to the UK public sector, while Crispin Robinson is Technical Director Cryptanalysis, GCHQ – the British government's head gamekeeper and head poacher respectively. Their paper sets out to argue that we do not face a binary choice between 'safe spaces for child abusers' and 'insecurity for all by default'. It sets out to provide more information about online crimes involving sexual violence against children in the hope of a more informed debate around public safety and exceptional access to encrypted messaging; we will provide still more here.

Levy and Robinson analyse the problem of online child sex abuse under seven categories. They separate communications between offenders into three categories, treating messages with indecent images separately, as well as group communications. Hunting for such offenders is no different from other signals intelligence tasks, except that the detection of historical illegal images may suggest new targets. The other four categories are:

- Grooming of young people by offenders, including sextortion – according to the US Department of Justice, the most important and fastest-growing cyberthreat to children, with 'more victims per offender than all other child sexual exploitation offences'[5];
- Online streaming of abuse, which typically involves young people in less well-policed countries being abused and the streams sold in Europe and the USA[6]. Here, effective policing typically involves the tracing of cryptocurrency transactions[7];
- Sexting ('consensual peer-to-peer image sharing') which although technically illegal in the UK, the USA and some other countries, is widespread among teens, and sometimes leads to sextortion;
- Viral image sharing, where naïve users share dubious images of children out of humour or disgust (these are generally not considered further).

---

[4] Commission of the European Union, "Technical solutions to detect child sexual abuse in end-to-end encrypted communications: draft document", Sep 2020, https://www.politico.eu/wp-content/uploads/2020/09/SKM_C45820090717470-1_new.pdf
[5] Loretta Lynch, "National Strategy for Child Exploitation Prevention and Interdiction", US DoJ, 2016
[6] Elie Bursztein, "Rethinking the Detection of Child Sexual Abuse Imagery on the Internet", Usenix 2019
[7] Stephanie Condon, "US, South Korea officials trace bitcoin transactions to take down massive child porn site", ZDNet, Oct 17 2019

This taxonomy is taken from the point of view of criminal and national-security investigators rather than from the more natural one of looking at harms to children. As such, it introduces an unhealthy bias. From the viewpoint of child protection and children's rights, we need to look at actual harms, and then at the practical priorities for policing and social work interventions that can minimise them.

**Evaluation and appraisal of policy options**

The role of scientists and civil servants is to provide advice by the evaluation of previous policy choices and the appraisal of new proposals. Both evaluation and appraisal should be driven by evidence. However, as Levy and Robinson admit, data on crimes of violence against children are hard to come by, whether online or offline[8]. We noted this in 2006 when writing a report for the ICO on the safety of children's databases[9] and again in 2009 when we followed up with "Database State", which became Lib Dem policy[10]. The 2010 election led to a coalition government and "Database State" led to the Munro review, during which we learned that the Department for Education routinely destroyed all serious case reviews – the reviews done when a child known to social workers is unlawfully killed. This was justified as "for data protection" but was patently about minimising the risk of embarrassment to officials and ministers.

Violent crime against children is largely family violence; worldwide, there are about 100,000 homicides of children a year[11]. The typical perpetrator is the mother's partner, although from puberty a growing number of homicides are committed by acquaintances. The 206 UK child victims of murder in 2020–21 represented a small increase on the usual toll of 170–180; this may be a side-effect of covid lockdowns. Child homicides are the visible tip of a largely invisible iceberg of child abuse, of which by far the most common kind is simple neglect. This is associated with multiple deprivation: with unstable families living in slums, structural unemployment leading to endemic poverty, despair leading to alcohol and drug abuse, and gangs providing role models. It should surprise no-one that patterns for sexual violence are similar, although different countries collect data in different ways (or not at all).

Survey data relevant to sexual violence with an online component include a 2017 analysis by the Canadian Centre for Child Protection documenting the experiences of 150 victims whose abuse was recorded and uploaded to the Internet[12]. Notable facts include that most victims were female (85%); abuse started for most victims (87%) before the age of 12; and where there were multiple offenders, the primary abuser was predominantly a family member (82%), with the abuse taking place in the family home. Where there was a single offender, he was still a family member half the time, while other offenders were typically local and in a position of trust, such as clergy, police, doctors or teachers[13]. This was followed by peer-to-peer abuse of and by older children. All these offence types may have an online aspect in that devices are used for surveillance and control of victims[14]. Violence that was initiated online rather than by someone already known to the victim makes up a minority of offences, even if we consider only those offences that resulted in harms visible online as well as contact abuse.

---

[8] For a survey of UK datasets, see Kairika Karsna and Liz Kelly, "The scale and nature of child sexual abuse: review of evidence", csacentre.org.uk, 2021; they note 'no estimate of the prevalence of online CSA' (p7)
[9] Ross Anderson, Ian Brown, Richard Clayton, Terri Dowty, Douwe Korff and Eileen Munro, "Children's Databases – Safety and Privacy", Information Commissioner's Office, November 2006
[10] Ross Anderson, Ian Brown, Terri Dowty, William Heath, Philip Inglesant and Angela Sasse, "Database State", Joseph Rowntree Reform Trust, March 2009
[11] Heidi Stöckl, Bianca Dekel, Alison Morris-Gehring, Charlotte Watts, Naeemah Abrahams, "Child homicide perpetrators worldwide: a systematic review", BMJ Paediatrics Open (2017)
[12] Canadian Centre for Child Protection, "Survivors' Survey – Full Report", at https://protectchildren.ca/en/resources-research/survivors-survey-results/
[13] Independent Inquiry into Child Sexual Abuse, "The Internet Investigation Report", March 2020

The media rhetoric around child sex abuse is unhelpful. In recent years it has become associated with far-right politics and its associated conspiracy theories, such as the Washington pizzeria allegedly used by Hillary Clinton to groom and traffic children for sex. Two tropes from these narratives surface in Levy and Robinson on p21: first, the claim that 2% of men are sexually interested in children, which does not square with their later claim that child sex abuse accounts are "orders of magnitude" below the 5% of abusive accounts of Facebook[15]; and second, the 'pathways to crime' claim that exposure to images of child sex abuse drives men to indulge in it. This mirrors the old claim that gay porn makes men gay, and the 'porn causes sex crime' view of social conservatives. No evidence is offered[16].

The evidence around images of abuse is also problematic. The bodies to which tech firms report abuse, NCMEC in the USA and the IWF in the UK, are membership organisations whose revenue comes from persuading more firms to subscribe, and their member firms have an incentive to report everything, however marginal, as a shield from liability. Levy and Robinson admit that last year's figure of 29.4m reports to NCMEC is unhelpful. They clarify that there were only 102,842 reports by NCMEC to the NCA, of which 20,038 were referred to local police, leading to over 6,500 suspects arrested or interviewed under caution and over 8,500 children 'safeguarded'. Statistics for prosecutions or convictions are not provided. If we estimate that the NCA's pipeline led to 750 prosecutions for indecent images[17], that was well under 3% of the 2019 total of 27,233 prosecutions for indecent image offences, of which 26,124 involved images of children[18]. What's more, the number of prosecutions peaked in 2016 and has fallen significantly since[19].

In short, the data do not support claims of large-scale growing harm that is initiated online and that is preventable by image scanning. Yet real harm has been done by false positives. The first wave of prosecutions for illegal abuse images, Operation Ore, swept up many innocent men who were simply victims of credit card fraud, their cards having been used to pay for illegal material. A number were wrongly convicted, and at least one innocent man killed himself[20]. The organisation responsible, CEOP, became part of SOCA and then of the NCA. It still uses as a metric the number of children 'safeguarded'. This term is elastic; it can mean that a child has been taken into care (whether rightly or wrongly)[21]; it can mean that a parent or carer has been arrested, or accepted a caution, or signed the Sex Offenders' Register, or (in the context of 'Prevent', which we discuss later) that society has been safeguarded from a child considered to be dangerous. The police have since acknowledged that too much effort has been put into indecent images and not enough into preventing actual abuse of minors[22].

---

[14] Karen Levy and Bruce Schneier, "Privacy Threats in Intimate Relationships", Journal of Cybersecurity 6: 1-13 (2020)

[15] Thanks to the Budapest Convention, the word 'child' in the CSA context can mean everyone under 18. The number of men who once fancied a 16-year-old when they were 18, and the number of adult men who fancy 6-year-olds, are very different. Details matter, and expansive claims must be treated with suspicion.

[16] The relevant evidence is mixed, with different studies suggesting that porn and sexual violence are substitutes or complements Substitutes: see Todd Kendall, "Pornography, Rape and the Internet", Clemson University, 2007; Complements: see Manudeep Bhuller, Tarjei Havnes, Edwin Leuven, Magne Mogstad, "Broadband Internet: An Information Superhighway to Sex Crime?" SSRN 1881507, 2011

[17] Overall some 12% of all CSA investigations resulted in a charge, down from 37% six years earlier

[18] There were also 6,387 CSA prosecutions leading to 4,870 convictions of which 2,211 involved images, so we could probably add about 3000 prosecutions to the denominator for image offences if we had the data

[19] Kairika Karsna and Liz Kelly, op. cit., p8, p53, p65 and p66

[20] D Campbell, "Operation Ore Exposed", PC Pro, July 2005; "Sex, Lies and the Missing Videotape", PC Pro, April 2007

[21] Where a child is taken into care that counts as a 'protect action'; we get no separate statistics for that.

Our context should therefore be crimes of sexual violence against minors. The European Parliament decided in 2017 not to use the American term "child pornography" but "child sexual abuse material" (CSAM) instead[23]. Rather than getting into a terminological dispute, we will use the term "CSAM" in what follows, but we prefer to read it as "Crimes of Sexual violence Against Minors". Germany considers children to be those under 14, while Denmark considers sex with minors under 15 to be a serious offence; yet the CSA Regulation will apply to images not just of children but of young people under 18. The term 'minor' is thus more accurate. Similarly, when we come to discuss terrorism, we will prefer the more general and less emotional terms "violent political extremism" for the kinetic variety and "violent online political extremism" for its online aspects.

**Local language models**

A novel and very problematic proposal is "to have language models running entirely locally on the client to detect language associated with grooming", in Levy and Robinson's words. This has since appeared in the draft UK and EU laws. Law-enforcement agencies largely gave up scanning emails for keywords such as 'bomb' 25 years ago as it doesn't work; the analysis of traffic data is much more effective, for which access to content is not essential[24]. The use of modern NLP models to detect illegal speech – whether sexual grooming, terrorist recruitment or hate speech – is highly error-prone. Our research group has long experience of looking for violent online political extremism, as well as fraud and spam. Going by text content alone, it can be difficult to get error rates significantly below 5–10%, depending on the nature of the material being searched for[25].

Modern messaging systems operate at such scale that filters need a false-positive error rate of 0.01% to be deployable, and 0.001% to be effective. Given the 10bn messages sent and received every day in the EU, even 0.001% would still mean 100,000 messages sent for moderation every day. The CSA Regulation envisages a new law-enforcement agency, modelled on and co-located with Europol, to deal with this traffic. Levy and Robinson note that the UK National Crime Agency triages 100,000 alerts a year from NCMEC, and that this takes 200 staff.

With 5% false positives, the task would not be feasible at all. Indeed, the European Commission assumed in an internal discussion document presented to the Council in June that they might get 10% false positives but then calculated that these would be 10% of the 1,000,000 true positives. However, the Commission got their arithmetic wrong. The false

---

[22] The Internet Investigation Report, op. cit., para 38 p 63
[23] Conclusion 4, 2015/2129(INI)
[24] Ross Anderson, "Crypto in Europe – Markets, Law and Policy", Cryptography – Policy and Algorithms 1995
[25] Anh V. Vu, Lydia Wilson, Yi Ting Chua, Ilia Shumailov, Ross Anderson, "ExtremeBB: Enabling Large-Scale Research into Extremism, the Manosphere and Their Correlation by Online Forum Data", arxiv 2111.04479, Nov 21

alarms would be 10% of all the texts processed, or a billion alarms a day[26]. Europe's 1.6m police officers would each have to triage 625 of them, which is hardly practical.

Modern spam filters, like filters for hate speech in social-media platforms, use many more signals than just message context. Engineers working on filtering for social-media firms agree that using content alone would significantly increase both the false positive and false negative error rates of filtering compared with using metadata as well. Levy noted the high error rates of ML-based content filtering elsewhere[27] and this is echoed in the later section in Levy and Robinson on 'Why AI and Machine Learning are not the answer'.

The Levy-Robinson paper cites research by Patrick Brours and Halvor Kulsrud but they found that 26–161 messages of a conversation were needed to sound an alarm on average[28]. Questions arise about its ecological validity, as their models are not trained on real child grooming data. Instead, they use two datasets, one of adult sexual conversations versus normal conversations, and one of adults pretending to groom other adults who are pretending to be child victims. Their classifiers ended up looking for words and bigrams such as "sweetie" and "lil slut". While their filters may "work" better than those of other academics using similar training data, there is no evidence they would be effective on real-world traffic. Levy and Robinson acknowledge 'significant positives', suggesting referral to a moderator at p28, yet screening 1bn messages a day in the UK would result in millions of false positives. There appears to be one trial product: Microsoft's Project Artemis has resulted in a chat scanning tool now available to system operators via Thorn, a private U.S. company with ambitions to be a key provider of scanning services. But there is no independent assessment of it, and Microsoft's claims on its behalf are notably weak: "This technique is not a complete solution, but it is a major step forward in becoming proactive in this fight."

It is hard to see how anyone could trust an NLP text scanning tool that was trained on data to which the public and even public-interest technologists have no access. There are too many ways in which machine-learning pipelines can be subverted: manipulating the inference engine, or the training data, or its labelling, or even its batching[29]. Levy and Robinson assume that NCMEC will be the curator of the dataset used to train the grooming classifier, but who will curate the terrorism recruitment dataset? And how can all the other actors in the supply chain be audited? We discussed these vulnerabilities and assurance issues in greater detail in 'Bugs in our pockets'[30]. The scanning of text makes more sense in the context of adaptive searching, where an investigator at a law-enforcement or intelligence agency could refine their search progressively until the output is small enough to be useful.

Dynamic searching by agencies operating in secrecy would not be compatible with audit and public accountability. In fact, it is entirely unclear how mandatory remote scanning of text on user devices could ever be squared with the human right to privacy and the established ban on mass surveillance without warrant or suspicion. With historical abuse images, there is at least the argument that a specific crime is being targeted. But if we have NLP models trying

---

[26] Sebastian Meineck, Chris Köver, Andre, "Interne Dokumente zeigen, wie gespalten die EU-Staaten sind", Netzpolitik, 12 Sep 2022
[27] Edinburgh, March 30th – see https://www.youtube.com/watch?v=qv6SS5FhdUk
[28] Patrick Brours, Halvor Kulsrud, "Detection of Online Grooming in Cyber Conversation", preprint
[29] E.g., Ilia Shumailov, Zakhar Shumaylov, Dmitry Kazhdan, Yiren Zhao, Nicolas Papernot, Murat A. Erdogdu, Ross Anderson, "Manipulating SGD with Data Ordering Attacks", arXiv:2104.09667 (2021).
[30] Hal Abelson, Ross Anderson, Steven M. Bellovin, Josh Benaloh, Matt Blaze, Jon Callas, Whitfield Diffie, Susan Landau, Peter G. Neumann, Ronald L. Rivest, Jeffrey I. Schiller, Bruce Schneier, Vanessa Teague, Carmela Troncoso, "Bugs in our Pockets: The Risks of Client-Side Scanning", Oct 14, 2021

to detect prohibited speech (whether for grooming or for terrorist recruitment) and throwing the product up to human moderators to screen for false positives – as suggested by the European Commission – we would have embraced full-blown bulk intercept and mass surveillance, which is contrary to established human-rights law.

At pp 57–58, Levy and Robinson suggest that an alarm could nudge the user to report the conversation to a moderator. An on-device abuse alarm under the user's control might well be lawful from the privacy viewpoint. But this makes a major assumption: that a moderator can be contacted. At present, despite the existing server-side scanning by social media, cases of sextortion are more likely to surface from user complaints. But their handling often leaves much be desired. It costs a service firm money when a human moderator intervenes in a conversation, and so they all try to limit this cost in various ways.

**Warnings and complaints from users**

Escape and reporting mechanisms are a vital safety feature for young users (and for other users too). Users who are pressured or harassed may need a quick-exit button, or to block someone who is adversarial or threatening, or to complain to the system operator, or to take a forensic recording of a conversation. What's more, the organisations tasked with dealing with CSAM agree that public reporting is the main way we identify new or previously unreported abuse material, and needs to be improved[31]. But the quality of the tools available for this purpose varies very widely across social media platforms, games and other online environments. Devising better mechanisms is an active area of research in our group.

My colleagues have analysed over 100 quick exit buttons they found on over 500 websites where grooming and harassment might take place; they also been studying whether abuse victims should be able to contact the police via chat rather than by dialling 999 or 111 (this is the subject of a project by the Suffolk police)[32]. Levy and Robinson, by comparison, point only at a Canadian guide to mechanisms for reporting child sex abuse material[33], not to mechanisms that would enable a child, a young person or even an adult user to seek help, to contact moderators, or to preserve messages for evidence. The more quickly people can report actual or attempted abuse, the better. And as Baroness Butler-Sloss famously warned, it is vital to treat each child 'as a person not as an object of concern'[34]. Helping children develop their agency to push back and protect themselves is part of helping them become independent adults.

A related matter is that social media firms fail to take threats against women seriously. Family violence typically involves assaults on women as well as on children, and most terrorist murderers first commit violent crimes against female family members (a point to which we will return later). It has been a standing complaint by women in the public eye since the mid-2010s, including female MPs, academics and others, that they receive floods of misogynistic abuse, including rape threats and death threats, about which neither the media companies nor in most cases the police are prepared to take action. One of the few welcome

---

[31] Denton Howard, "INHOPE feedback on the EC proposed regulation laying down rules to prevent and combat child sexual abuse", 25 September 2022
[32] Kieron Ivy Turk and Alice Hutchings, "Click Here to Exit: An Evaluation of Quick Exit Buttons", work in progress.
[33] Canadian Centre for Child Protection, "Reviewing Child Sexual Abuse Material Reporting Functions on Popular Platforms", https://www.protectchildren.ca/en/resources-research/csam-reporting-platforms/
[34] Lord Justice Butler-Sloss, "Report of the Inquiry into Child Abuse in Cleveland", 1987

features of the Online Safety Bill currently before the UK parliament is that it will enable Ofcom to impose duties of contactability on regulated service providers[35]; the EU's Digital Services Act also sets out to improve the removal of illegal content. However, the provisions are flabby, and the law could do more. At present, tech firms pay attention to takedown requests from the police and from copyright lawyers, as ignoring them can be expensive – but ignore ordinary users including women and children. That needs to be fixed, whether by criminal sanctions or by significant financial penalties[36]. One may expect the tech firms to lobby hard to mitigate the cost and liability, but this may be one issue on which child-safety and privacy campaigners can work together to ensure that safety mechanisms based on user reporting are robust. Regulators will need not just the powers but the political backing to force big service firms to respond quickly to complaints, to take down illegal content across a wide range of harms from sexual offences through death threats to fraud, and to keep it down.

**The role of machine learning**

The Levy-Robinson paper lacks clarity in its discussion of AI. They make various claims about the capabilities of machine learning, which appear somewhat confused. They seem to be sometimes talking of AI on content and at other times of AI on metadata. Some of the criticisms apply whether the algorithms used for abuse detection are modern deep neural networks (DNNs) or old-fashioned regressions; for example, they assert without any evidential support that victim-generated reporting will lead to highly biased data (p52). Yet survey evidence indicates that half of victims of serious sextortion offences try to report them to the platform, and 89% to friends or family[37]. The current policy of searching for 'previously known content' already leads to biased patterns of arrests.

Levy and Robinson are also defeatist about approaches other than spotting historical images in the face of 'adversary behaviour changes'. Yet the IWF claims that, thanks to its efforts, the number of abuse websites hosted in Britain has fallen by two orders of magnitude. Adaptive behaviour is inevitable in adversarial settings. An optimist might hope that the vetting and barring regime is cutting abuse by authority figures such as priests, teachers and police officers[38]. The least adaptive forms of abusive behaviour may well be two of the most persistent and serious categories. The first is the incestuous families: men who were abused as children are more likely to abuse their own families in turn. The second is the pimps, who tend to exploit young women who have been neglected in dysfunctional families or in care settings, and who are desperate for anything resembling affection. This strategy works online as well as offline, and the sex trade has been fairly resilient in the face of many governments' attempts to regulate and control it.

Levy and Robinson criticise Facebook's AI for having much higher error rates than PhotoDNA, but this is an unfair comparison as PhotoDNA looks for known images while the Facebook system in question looks for new material. The realistic comparison would be with the ML systems they propose for text scanning. It has been reported that Facebook caught

---

[35] Ofcom, "First phase of online safety regulation – Call for evidence", 6 July 2022
[36] Ross Anderson, Sam Gilbert, Diane Coyle, "The Online Safety Bill", Bennett Institute, October 2022
[37] Wendy A Walsh and Dafna Tener, ' "If you don't send me five other pictures I am going to post the photo online": a qualitative analysis of experiences of survivors of sextortion', Journal of Child Sexual Abuse (2022) v 31 no 4 pp 447–465
[38] It would still be useful to see comparative research on abuse patterns between the UK with its heavyweight vetting and barring scheme, and countries like Germany where kids are still 'free range'

only about a quarter of hate speech in English and more like 2% in Arabic[39]; yet in the West Bank, its filtering in Arabic has been become so heavy-handed that Palestinian Arabs use an old font to circumvent it[40]. Given that endpoint scanning would have even more false positives, the Facebook experience is not good news for the proposed text-scanning policy. If text scanning ends up requiring collaboration with the network / server on metadata, where the client NLP acts somewhat like a smart speaker listening for a 'wake word' – or even ends up supporting adaptive searching by investigators – then this takes us directly to mass surveillance, with no prospect of compliance with human-rights law.

Levy and Robinson also ask how a deep neural network can drive warrants. From industry's viewpoint, the answer is simple enough. An indication from their NLP scanner that '*with probability 60% a child is being groomed*' will pass the civil threshold for tort liability if the platform fails to act. So the firm will feel they have to do something. In a world without cost constraints, they could call the putative victim and talk to them. If a victim agrees that attempted grooming may have taken place, the investigator can work with them to invoke forensic preservation, or ask for the backups[41]. But human intervention costs money; and given the current state of the law and of industry practice, it's more likely that the suspect (and perhaps their family and friends) will have their accounts blocked.

Levy and Robinson's claim that false positives from machine-learning classifiers can be dealt with by human moderation is simply wrong. The agencies have little liability for the human cost of false positives, and neither do the tech companies. In a recent and widely reported case, a father took a picture of his toddler son's genitalia, which had become swollen, at the request of a nurse. He sent them to their doctor, who duly prescribed antibiotics. The image was uploaded automatically by his phone to Google, which classified it as abuse and suspended his accounts. This led also to a visit by the police, who cleared him; yet he has been unable to get his accounts back[42]. This case raises not just issues of privacy rights (of both the parent and child) but also whether services with market power should have a universal service duty. There is a serious confounding factor, namely that indecent images of children are a strict-liability offence in both the USA and the UK. This makes sensible action by many potential guardians, from service firms to parents, teachers and social workers, much more difficult. We will return to this problem later.

Levy and Robinson also claim that machine-generated approximate labels won't help. This may well be true for sextortion, a subject we need to consider in more detail.

**Policy options for the prevention of sexual aggression**

Crime prevention can be primary, secondary or tertiary: we can seek to remove the social cause of crime, make specific crimes harder, or lock up and rehabilitate offenders[43]. In the case of the broad problem of crimes of violence against women and children, scholars

---

[39] Isabel Debre, Fares Akram, "Facebook's language gaps weaken screening of hate, terrorism", AP, Oct 25 2021
[40] Mustafa Abu Sneineh, "Facebook users deploy old Arabic font to bypass algorithm, support Palestinians", Middle East Eye, 23 May 2021
[41] Note that only Android phones support end-to-end encrypted backup; iPhones, which are much more popular among teens, do not, and so backups can be retrieved from the iCloud with a warrant.
[42] Kashmir Hill, "A Dad Took Photos of His Naked Toddler for the Doctor. Google Flagged Him as a Criminal." New York Times, August 21 2022
[43] Paul Brantingham, Frederic Faust, "A Conceptual Model of Crime Prevention", Crime and Delinquency 1976

identify two types of cause: feminist scholars identify a cultural cause in misogyny, while the mainstream view emphasises a socio-economic cause in poverty and deprivation. We will return to misogyny later. On the mainstream view, the best primary prevention option would be to increase child benefit while the best secondary prevention option would be to increase the number of child social workers. One recent improvement is that sex education has been mandatory in schools since 2020; this means teachers can discuss sexual offences with the whole class, and empower minors to understand, detect and resist predatory behaviour.

In the specific case of sextortion, both prevention and policing can be hard for the following reasons.

- About a third of teens send explicit images to each other as part of the normal process of flirting. Those who are under 18 are committing a crime in the UK or the USA, as the possession of indecent images of under-18s is a strict-liability offence. As Levy and Robinson note, the UK police and CPS will normally not prosecute (so-called 'Outcome 21') but this is not guaranteed if there is further dissemination. Strict liability has a chilling effect on the ability of teachers to deal appropriately with sexualised bullying, as we'll discuss later.
- When relationships break up, one of the participants may either disseminate an image of the other without consent ('revenge porn' / 'non-consensual intimate imagery', or NCII) or threaten to ('sextortion').
- In the former case, children who report NCII are admitting offences, and their images may end up on the watch list as IWF/NSPCC 'Report remove' (L&R p 16).
- In the latter case, the great majority of teen victims and offenders are peer-to-peer with about 5% of straight kids reporting victimisation and about double that for gay kids. The offenders are usually peers of the victim, with 2% of kids (almost all male) reporting offending[44]. About 10% of teens experience physical dating violence and about 15% experience some sexual victimisation.
- Some of the minor victims of sextortion are targeted by adult abusers who may try to meet them for sex, to draw them into prostitution, or even to bully them into sexually abusing other children around them[45]. In many of these cases, the victims are already vulnerable because of neglect; family circumstances may include drugs, alcohol and violence, and cared-for children are also at risk. The exploiters initially show interest and affection, and then get compromising images which are used for blackmail[46]. This is the modern version of the age-old modus operandi of the pimp – looking for kids who get no affection anywhere else. Such cases are rightly the focus of social-work and police intervention[47]. In extreme cases such exploitation can lead victims to commit suicide.
- Less serious cases must be dealt with by friends, parents or teachers, as there are too many of them for the police. Both sextortion and revenge porn are suffered by several percent of adults too; they both count as volume crime. 'The dark side of online

---

[44] Justin W. Patchin, Sameer Hinduja, "Sextortion Among Adolescents: Results From a National Survey of U.S. Youth", in Sexual Abuse, September 28, https://doi.org/10.1177/1079063218800469
[45] Roberta Liggett O'Malley and Karen M Holt, "Cyber Sextortion: An Exploratory Analysis of Different Perpetrators Engaging in a Similar Crime", Journal of Interpersonal Violence 2022, v 37 no 1–2 pp258–283
[46] L&R 2.3.4 short section with focus on CSAM being sent as part of grooming process, which happened but not in most cases... perps are more likely to sell the images or use them as a threat to silence the victim
[47] Benjamin Wittes, Cody Poplin, Quinta Jurecic and Clara Spera, "Sextortion: Cybersecurity, teenagers and remote sexual assault", Brookings, May 2016

dating', they are involved with romance scams and other cybercrimes[48]. Their incidence has risen during the pandemic, and is higher for both racial and sexual minorities[49]. The police do not currently prioritise such offences; many forces avoid investigating crimes that involve technology, or remote offenders.

Online content scanning alone cannot hope to fish out the serious cases of exploitation by pimps and other adult predators from the ocean of peer bullying at school and extortion of money from vulnerable adults. One of our informants who worked with a national police force remarked, "If a guy with tattoos turns up at the school gates in a tuned car, you know he's up to no good." Local know-how is key; defunding the vice squad and replacing it with analysts looking at screens is unrealistic. Another problem is that as soon as Google or Facebook knows that an account has indecent images of children, they currently have no choice but to close and lock it in the expectation of warrants (I explain the law below). If large numbers of accounts are referred for moderation – whether because of false positives, or because an ML tool starts picking up sexting that was previously ignored – then significant numbers of children and young people may suddenly lose access to their digital lives[50].

In cryptography policy, it is customary to discuss protocols in terms of Alice and Bob. Here, Alice and Bob are 16-year-olds at the same school who have a relationship. As the age of consent in the UK is 16, this is quite lawful. They break up. Bob makes an ambiguous remark about an intimate video of Alice, perhaps along the lines of "Gosh, and you're such a porn star!" Alice makes a hostile reply, and next day the video appears online. Alice's friends Carol, Dave and Eve remark on it. One of the frequent current outcomes is that Alice is so embarrassed she moves to a different school. This may affect her exam results, so she doesn't get into her chosen university course and perhaps misses out on her chosen career.

What's the rational crime prevention here? Good sex education so that, first, Alice doesn't make intimate videos and Bob doesn't push her; second, that if Bob does find himself with an intimate video of an ex-partner he's sufficiently aware of serious consequences that he doesn't share it, or threaten to. Finally, serious consequences – which in this case could mean expulsion from the school or police action depending on the circumstances[51]. There also needs to be a process for Alice to contact the platform (Facebook, Apple, Google) and have her video removed – a process that's at least as responsive as the mechanisms that remove material whose copyright belongs to commercial studios, and which also doesn't result in the loss of all her online accounts. The head teacher also needs to be able to get evidence. Both outcomes will require a change in policy.

The international law on child pornography stems from the Budapest Convention of 2004, of which article 9 mandates that signatories criminalise the production, making available and distribution of material that visually depicts a minor engaged in sexually explicit conduct, when committed intentionally and without right. UK law stems from the Protection of

---

[48] Cassandra Cross and Roberta Liggett O'Malley, ' "If U Dont Pay They will Share the Pics": Exploring Sextortion in the Context of Romance Fraud', Victims and Offenders, May 2022
[49] Asia A Eaton, Divya Ramjee and Jessica F Saunders, "The Relationship between Sextortion during COVID-19 and Pre-pandemic Intimate Partner Violence", Victims and Offenders, January 2022
[50] It also allows denial-of-service attacks that could be used for mischief, harassment or even blackmail.
[51] Students at UK universities typically have the option of using internal disciplinary procedures to deal with sexual harassment cases where the complainant decides not to make a complaint to the police. Often the preferred outcome for the victim is to no longer see their abuser in class, rather than a 5% chance of a conviction for indecent assault in three years' time. Teachers believe such options should be available to schools too, so adolescents can make mistakes without ending up on a pathway to prison. But the law can get in the way.

Children Act 1978 and the Criminal Justice Act 1988, under which intent is irrelevant, including the motive of the taker and the circumstances. There is a statutory defence of 'legitimate reason', but this is surrounded by uncertainty arising from case law. It also applies only to possessing indecent images, not to making them; for Alice to not be prosecuted for her nude selfie, the CPS must exercise its discretion and decide that prosecuting her would not be in the public interest.

This is why even the Internet Watch Foundation operates under a 'letter of comfort' from the Crown Prosecution Service setting out operational parameters within which it will not be prosecuted for collecting, collating and reporting abuse images. While a parent concerned about their child's health might simply take the picture and, if challenged, argue that it's not an indecent image but a medical image, the position of service firms is much more difficult. Corporate lawyers understandably demand the permanent removal of accounts where indecent images have been found or suspected[52]. The UK courts tend to the view that every time an image is copied, it is "made" afresh, which leaves firms with no 'legitimate reason' for showing a potential abuse image to anyone except the authorities.

The consequence of the UK's approach is that if Alice (or her mother) reports attempted extortion to the school, they may not feel able to do anything. If she reports it to the authorities, then the police or CPS may decide there isn't clear enough evidence of extortion. As it's an indecent image, it's unlawful for the school or her parents to possess it, and Alice has just confessed to making it, which is an offence under s1(1)(a) PCA 1978. The video may indeed end up on the IWF watchlist, but then her friends Carol, Dave and Eve may find themselves on the NCA's target list. By possessing indecent images and not reporting them at once to the police, they each committed a PCA offence. Even if no action is taken, the NCA report will be retained forever 'for intelligence purposes'[53], so Carol, Dave and Eve may later find themselves barred from occupations that involve a security clearance or contact with vulnerable people[54]. For all these reasons, it may be rational for Alice to decide it's not prudent to report Bob: the likelihood of a proportionate response against him is low, while the risk of secondary harm to herself, her friends or her parents is real.

In short, the UK's current approach is not fit for purpose, and the EU should not adopt it. Instead, careful thought must be given to subsidiarity. At which level in the enforcement hierarchy should abuse be dealt with? Digital solutionism favours centralisation as tech companies want a convenient single point of contact (NCMEC, NCA, Europol?) while effective child protection requires local engagement with multiple guardians (police, schools, social work, addiction services...). But local guardians need access to forensic data, and in a typical case of sexualised cyberbullying the traffic data will often be more relevant than image content. The complainant will already have the images; the question of interest to teachers, social workers and local police officers will be who sent messages to whom when. But if the accounts are closed by the service firms as soon as the images are brought to their attention, even the traffic data is unlikely to be available, except to the police – who only investigate the most serious cases.

The argument will still be made by the agencies that scanning for historical images catches some serious offenders, even though they can't quantify how many. Even though most abuse is in person, and most guardians are local, we accept that occasionally a man abusing his stepchild may be caught after images of other children are detected by server-side scanning. It may even be true that the number of such arrests might fall if Facebook Messenger moves to end-to-end encryption, although here the evidence seems extremely weak: the number of UK men arrested or interviewed for image offences seems to be constant at several hundred per month despite the proliferation of end-to-end encrypted messenger products since 2015.

---

[52] U.S. law also provides strict liability for indecent images though the details are different.
[53] Levy & Robinson, op. cit., p 13
[54] Sandra Paul, "Sexting: "Outcome 21" - a solution or part of the problem?", Kingsley Napley, May 4 2018

It is often argued that electronic surveillance helps law enforcement, but in those cases where we can measure its effects, the evidence often falls short. In one example that is being discussed in Germany in the context of the proposed CSA Regulation, the security and intelligence agencies had pushed for the EU Data Retention Directive, which came into force in 2008, but was later in 2010 found to be unlawful. In this natural experiment, bulk data that in theory enabled CSA offenders' IP addresses to be linked better to identifiable suspects were available to the German police for two years, but not before or after; and they were available to Swiss police before the German police got them, and even after the ECJ ruling. Meanwhile, other countries such as Sweden did not retain data at all. The statistics reveal that access to data did not have a deterrent effect, did not assist in clearing up crimes, and did not increase convictions[55]. So when governments argue that privacy must be impaired once more, in the hope of arresting more abusers, some scepticism is called for. But that is not all.

**Human rights and the rule of law**

When it comes to any discussion of means and ends, our starting point is the need to maintain a democracy based on the rule of law and founded on human rights. Levy and Robinson's casual discussion of a 'social contract' on p 41, based on a survey of service users saying that detecting child sexual abuse was as important as, or more important than privacy, is disturbing. The social contract involves all of society, not just the operator and users of a particular online service; and human rights cannot be waived just because some opinion poll suggests that people would like to do bad things to the villain of the day.

The justification of torture following 9/11 was based on just this kind of utilitarian populism, with the media repeatedly playing the ticking bomb argument. Again and again, human rights defenders were pressed by TV interviewers on whether, if the police believed that a jihadi suspect knew the unlock code to a ticking nuclear bomb in Manhattan, it would be morally justified to torture him. Fox even parlayed this into a TV series. The outcome was a disaster for NATO both militarily and politically. It led to Abu Ghraib, destroying NATO's moral standing. When NATO troops tortured Al-Qaida suspects in the same prison as Baathist secret police, they facilitated the emergence of ISIS, which caused mass casualties and multiple atrocities – from forced conversion to child marriage and slavery. The rule of law in the USA has been gravely compromised by the suspects who are detained for life at Guantanamo as they cannot receive a fair trial. The loss of moral standing has deepened cynicism and fuelled the growth of the anti-democratic movement worldwide.

The intelligence agencies supported that programme and continue to use its utilitarian arguments. Levy and Robinson's flippant use of the term 'social contract' suggests that the lessons have not really been learned. The relevant part of the social contract is the European Convention on Human Rights, as the UK remains within the jurisdiction of the European Court of Human Rights. The right not to be tortured remains absolute, and the right to privacy may be limited only insofar as is proportionate, necessary and in accordance with law. The European courts have repeatedly struck down surveillance measures that apply to all without warrant or suspicion, and the UK Investigatory Powers Act prohibits the use of bulk intercept against UK persons. U.S. law also requires surveillance to be targeted. The attraction of historical image scanning for the intelligence agencies is precisely the argument that it provides enough targeting to justify dragnet content scanning.

---

[55] Hans-Jörg Albrecht et al., "Schutzlücken durch Wegfall der Vorratsdatenspeicherung?" Max-Planck Institut für ausländisches und internationales Strafrecht, Freiburg, July 2011

Necessity and proportionality must be assessed with respect to some lawful policy objective. If the objective is to minimise crimes of sexual violence against minors, the relevant context is violent crime, as we discussed above. The fact that possession of indecent images of minors is a strict-liability offence may help intelligence agencies argue that 'all indecent images of minors' should be considered a legitimate target category. But it creates a significant law-norm gap and disempowers children's real guardians. The necessity and proportionality tests are failed. And as well as sextortion, guardians must also deal with revenge porn – a closely-related problem, but one that is largely disregarded by policymakers and is largely missing from the Levy-Robinson taxonomy[56].

**Policy options around terrorist radicalisation and recruitment**

Although Levy and Robinson's focus is child abuse, Britain's Online Safety Bill would also empower the authorities to impose server-side scanning of text by NLP programs for terrorist radicalisation and recruitment. The Metropolitan Police in London has run an Internet Referral Unit for ten years that looks for terrorist material online and asks service providers to take it down; since 2021 Europol in Brussels has been doing the same. Extending this service from material published online to private material would be a major change, and the history of function creep in surveillance suggests that client-side scanning would be the logical next step, if such scanning is permitted for child protection. Some words on terrorism are therefore in order.

A growing body of scholarship challenges the approaches to terrorism taken in the UK since 9/11. One aspect is the very strong link between misogyny and violent political extremism, which extends across both Islamist and far-right violence. As one striking example, Joan Smith has studied all the terrorist murders in Europe since 9/11 and a significant number of mass shootings in the USA[57]. In the great majority of cases, the killer committed a violent crime against a female family member before going on to murder members of the public. The scholar Val Hudson writes of the 'third day effect': on the first day of a mass shooting you hear about the killer, on the second day you hear about his victims and on the third day you hear about the female relative he beat up or killed first. Both terrorists and mass shooters were very often raised in households with a hyper-dominant and abusive father. However toxic masculinity affects not just families but whole societies. Val Hudson has also led the creation and curation of the Womanstats database which enables analysis between states of the relationship between misogynistic attitudes and practices and aspects of development such as economic growth and political stability. Tribalism comes in many variants, with women in different societies suffering the effects of child marriage, forced marriage, bride-price, dowry, female genital mutilation, claustration, polygyny and polyandry, as well as inequality in education, employment and life outcomes. In its various forms, tribalism and its associated syndromes of aggressive masculinity are associated with political instability, violence, and poor socioeconomic outcomes[58].

---

[56] The honourable exception is Vice-President Kamala Harris, who as Attorney-General of California worked to get tech companies to deal more expeditiously with NCII material. However, the most effective results involved the New York Times bringing pressure against VISA and MasterCard to cut off the revenue streams of Mindgeek, which operates Pornhub and other major porn sites. See Kari Paul, "Pornhub removes millions of videos after investigation finds child abuse content", The Guardian, 14 Dec 2020. The Levy-Robinson characterisation of Pornhub as "legal" ignores this.

[57] Joan Smith, "Home Grown: How Domestic Violence Turns Men Into Terrorists", 2020

Rather than using bulk intercept to look for young men who 'express an opinion or belief supportive of' al-Mohajiroun or download a copy of the Anarchist's Cookbook, the police need to pay more attention to violence against women and children. If there are two hundred youths hanging out in a mosque with a fiery Salafist preacher, the man to watch is not the one who downloaded a cookbook but the one who beat up his sister. Other factors do matter, from a violent father through mental illness to life crises, but violence against women appears to be the strongest signal. This once more means local police work, not centralised electronic surveillance. Much the same holds for extreme right-wing violence, which since 2018 has surpassed Islamist extremism in terms of reported offences[59]. Our own research into violent online political extremism has confirmed strong links with the manosphere and misogynistic hate groups online[60]. Unfortunately, this modern scholarship on violence is mostly neglected by the Five Eyes' law-enforcement, intelligence and foreign-policy establishments[61].

The current government strategy on terrorism, Prevent, is highly polarising. Conceived after the 7/7 London bombings, it initially meant providing money to civil-society organisations like youth clubs to spy on young people for the police; after the 2010 election the duty shifted to teachers, GPs, social workers and other public-sector staff. Enrolling them as police agents created a conflict of interest with their basic preventative role of supporting and nurturing kids. Prevent was also seen as a means of spying on Muslim communities, and the Antiterrorism Crime and Security Act 2016 extended its scope to all extremism, rather than just violent extremism. The UK is now the only country in the world to impose a duty on its health service to report radicalisation[62]. These mechanisms still disproportionately target Muslims, despite the fact that the extreme right has committed most UK terrorist offences since 2018–9. The Home Office refuses to talk to Prevent Watch, a group of academics who conducted an independent review; it has instead appointed an 'independent reviewer' of Prevent – who has been criticised for being a former member of a right-wing think-tank and for having appeared to support torture[63].

Prevent has also been widely criticised by mainstream criminologists. Quite apart from concerns about effectiveness, there are serious ethical issues with a strategy of 'disruption' where police powers are used to harass people who are thought might commit crimes in the future or even just 'radicalise' others – without any independent oversight or adversarial

---

[58] Valerie Hudson, Donna Lee Bowen and Perpetua Lynne Nielsen, "The First Political Order: How Sex Shapes Governance and National Security Worldwide", Columbia, 2020
[59] See figure 6 of "Individuals referred to and supported through the Prevent Programme, England and Wales, April 2020 to March 2021", UK Home Office, 18 Nov 2021
[60] Anh V. Vu, Lydia Wilson, Yi Ting Chua, Ilia Shumailov, Ross Anderson, "ExtremeBB: Enabling Large-Scale Research into Extremism, the Manosphere and Their Correlation by Online Forum Data", arXiv:2111.04479 (Nov 8 2021)
[61] With the honourable exception of Hillary Clinton when she was US Secretary of State.
[62] Adrian James, "I'm a doctor, not a counter-terrorism operative. Let me do my job", The Guardian, Mar 21 2018
[63] Fahid Qurashi, "The Last Stand: Shawcross and the Prevent Review", Cage, Feb 5 2021

judicial testing of the evidence[64]. Despite this controversy, its 'pre-crime' methods have been exported to other policing areas such as cybercrime, and to other countries. Governments anxious to spend money on countering violent extremism have incentivised many business, legal and bureaucratic entrepreneurs to come up with de-radicalisation schemes; ministers generally lack the data to tell what's working and what isn't, and the political courage to cut off those initiatives that are manifestly not working[65].

Defining extremism is hard, as it can be anything from the Daily Mail line on Europe to violent neo-Nazis. In this context it is disturbing that the Online Safety Bill offers an exception for journalism that covers established news media but seems to exclude many others who perform a watchdog function by holding power to account – such as bloggers. The wide boundaries we set for print media should apply to online media too, giving a single standard for free speech, rather than trying to micromanage online censorship for the benefit of the government of the day. We should draw the red line at inciting violence, and we also need a grown-up discussion about cruelty, and about policing threats of violence. Violent online political extremism is a problem, but it's the violence we need to focus on. Our current definitions of online hate speech are problematic; some arrests get entangled in the culture wars[66]. Legal scholars argue that speech offences must be defined using narrow and tested concepts such as libel and incitement, rather than broad-brush ones such as hate and care[67]. Searching text for 'extremism' could mean searching for just about anything that the government of the day dislikes. The prospect of getting wide public support for such searches is close to zero. Censoring 'extremism' would thus feed the narratives on which extremism relies – that the UK government doesn't care about its citizens, that it spies on them, that it considers itself above the law, that officials lie and ministers are corrupt, and so on.

Finally, when dealing with terrorism, it is counterproductive to talk it up and use it as an excuse. In the words of Salman Rushdie[68], "How to defeat terrorism? Don't be terrorised. Don't let fear rule your life. Even if you are scared."

**Practical child protection**

Rather than using kids and terrorists as an excuse to expand bulk intercept capabilities, governments need to calmly revisit several policy areas, including family violence, political violence, and online crime. Details matter; they will vary from one country to another depending on local law, police practice, the organisation of social work, the availability of firearms, and political polarisation (this list is not exhaustive).

In what follows I focus on the UK. On child protection, the primary crime prevention strategy is to reduce the number of families living in poverty. The secondary crime prevention strategy is to increase the number of capable guardians, such as child social workers, and empower them in various ways. Tertiary crime prevention must involve more attention to reports of violence against women and children. This is also critical to mitigating violent political extremism, as perpetrators of terrorist violence usually started with violence against female family members and have often themselves been victims of violence when younger. It also involves the rehabilitation of offenders, from which quite a lot can be learned[69]. (Primary prevention of political violence is by its nature political; so please don't feed the memes of the global anti-democracy movement, such as by blaming international liberal elites or banging on about grooming and paedophilia.)

---

[64] Martin Innes, Colin Roberts and Trudy Lowe, 'A Disruptive Influence? "Prevent-ing" Problems and Countering Violent Extremism Policy in Practice', Law and Society Review v 51 no 2 pp 252–281 (2017)
[65] Lydia Wilson, "For 20 years we've tried to stop extremists. Instead, we've created a wasteful industry", New Lines, September 10 2021
[66] Dan Sales, "Fury over arrest of Catholic mother over claims she posted 'malicious' messages online: Campaigners condemn Surrey Police for 'wasting time on trivial nonsense' after swooping on vicar's wife following 'Twitter spat about gender issues' ", Daily Mail, 5 Octoer 2022
[67] Graham Smith, "Reimagining the Online Safety Bill", Cyberleagle, 18 August 2022
[68] Salman Rushdie, 'Let's get back to life', The Guardian Oct 6 2001

Online crime in general has since 2005 been largely ignored by our law enforcement and intelligence agencies, a matter on which parliamentary committees have commented repeatedly. Thankfully, the previous Home Secretary announced a new focus on cybercrime with the appointment of a new NCA director[70]. Romance scams and their associated sextortion against adults have for years been just as low a priority as crypto crime and authorised push payments. It is high time for online sexual extortion – including NCII, and crimes against adults – to move up the agenda.

However, the drive by GCHQ and the NCA for centralised techno-solutionism is the wrong way to tackle complex social problems. The local guardians who do most of the work, such as social workers, GPs, schools and police, have been starved of funds for years. Proper protection also means empowering children and young people themselves, which in turn requires respecting their rights. This does not mean undermining encryption; nor does it mean introducing a generalised monitoring obligation, a point on which the Online Services Bill is suspiciously ambiguous[71]. It means making it much easier for kids (and other service users) to report illegal material, harassment and abuse online. It also means that services that fail to take illegal material down must face fines that are large enough to hurt. That is the real opportunity for the Online Safety Bill: using the law to empower users and local guardians, rather than to empower central agencies. The success of France and Germany in getting the tech majors to block Nazi material and symbols in their jurisdiction is worth study. Swastikas may not be illegal in Britain, but lots of material is – and not just indecent images. (It's also worth noting that the slowest of the large firms at taking down abuse is not currently Facebook – GCHQ's target – but Twitter[72].)

It is welcome that the government is hiring more police officers, many of whom need to be directed to work on family violence. But teachers, social workers and parents also need to be able to exercise their proper role as guardians. The strict-liability prohibition of indecent images of minors gets in the way of child protection work in schools and elsewhere in the community, while creating a perverse incentive for national intelligence agencies to engage in issues of child protection in which they have little expertise and for whose failures they are unlikely to be held accountable. The Protection of Children Act 1978 should be amended so that indecent-image offences require intent, like almost all other criminal offences, and in line with the Budapest Convention. If Parliament cannot find the courage to do that, then as an interim measure the CPS needs to enable teachers and social workers to deal with sexualised bullying which may involve some images of minors by giving them an appropriate liability shield, just like the IWF. It also needs to enable the service providers to take a less risk-averse approach to the scanning they already do, so they do not end up penalising innocent users who suffer false positives. For both purposes, the CPS must clarify the interpretation of "legitimate reason" for possession and the public-interest test where possibly indecent images are 'made'[73]. At present, the application of the public-interest test appears to be a decision for local police forces[74]; no corporate lawyer will want to rely on the hope that every Chief Constable will always be a paragon of wisdom, mercy and beneficence.

---

[69] Rebecca Myers, "Inside Job: Treating Murderers and Sex Offenders. The Life of a Prison Psychologist," HarperCollins 2022
[70] Caroline Davies, "Graeme Biggar appointed director general of National Crime Agency", The Guardian, Aug 12 2022
[71] Heather Burns, "Fixing the UK's Online Safety Bill, part 1: we need answers", webdevlaw.uk, 31 July 20
[72] Zoe Schiffer and Casey Newton, "How Twitter's child porn problem ruined its plans for an OnlyFans competitor", The Verge, Aug 30 2022

Institutional reform is also overdue, as many police forces are not much good at dealing with crimes against women and children. Non-UK readers may not be aware, but London's mayor fired its police chief after an officer arrested, raped and murdered a young woman, and women protesting about this were beaten up by other officers. It was then discovered that 170 known or suspected abusers had not been dismissed or suspended from the force[75]. The then UK Home Secretary rightly demanded radical reform from the incoming chief[76].

The measurement of police performance is complex, and real change requires programmes that are funded and carried through with resolve over many years. Luckily we have a few examples of progress, for example against drug dependency and homicides[77]. Local police forces need to give a higher priority to family violence; they also need both the capacity and the incentive to deal with volume crimes that have a technology component. That in turn will require changes to how police forces and the NCA interact.

Neither local nor central agencies have effective ways of dealing with false positives under the current system of server-side scanning, and until that's fixed no-one should even think about expanding the programme. We mentioned Operation Ore; the men wrongly convicted there must have their cases reopened. Even the U.S. military has now acknowledged that it needs accountability to get institutional learning, and mechanisms for countering the confirmation bias endemic to large organisations; it has therefore reformed its procedures for investigating and minimising civilian casualties[78].

The Department for Education, which supposedly leads on child protection, had better stop destroying records of serious case reviews and start trying to learn from them. If not, it deserves to lose its lead on the issue; but which department should lead instead? Not, most people would think, the security and intelligence agencies, as they have historically faced no sanction for failing to prevent harm. GCHQ's excesses, from the Five Eyes' "collect it all" strategy to its hoarding of intimate Yahoo video chats of innocent citizens, were disclosed by Edward Snowden. The security service, which would presumably want to curate the data used to train ML models for detecting radicalisation, has recently suffered the exposure of its long record of xenophobia and torture in wars from Ireland through Iraq, Palestine, Kenya and Malaya to Northern Ireland[79], and has had its retention of data obtained under warrants declared 'undoubtedly unlawful' by the Investigatory Powers Commission[80].

---

[73] Or ministers must, whether by regulation or (best of all) by amending POCA 1978.
[74] Levy & Robinson, op. cit. 2022, p 14
[75] Rachael Burford, "More than 170 serving Met Police officers under investigation for alleged domestic abuse", Evening Standard, Sep 1 2022
[76] James Gregory, "Metropolitan Police must learn from appalling mistakes – Patel", BBC News, Sep 4 2022
[77] Richard Lewis, "I'm a police chief – Liz Truss's soundbite policies won't help us fight crime", The Guardian, Sep 2 2022
[78] Eric Schmitt, Charlie Savage and Azmat Khan, "Austin Orders Overhaul to Better Protect Civilians During U.S. Combat Operations", New York Times, Aug 25, 2022
[79] Caroline Elkins, "Legacy of Violence: A History of the British Empire", Penguin, 2022
[80] "MI5's use of personal data was 'unlawful', says watchdog", BBC News, 11 June 2019

With crimes of violence – whether crimes of political violence, crimes of violence against children, or even crimes that fall into both categories – the policing priority must be the primary offences. For police officers, social workers and others involved, protection work is hard and thankless, and people may easily be distracted. Rather than working with desperate people in unpleasant slums, it may be tempting to sit at a computer and play with machine-learning models, or even buy Google ads telling teens to 'just say no' to grooming. Such displacement activities should not be encouraged. Good policy involves ministers coordinating multiple actors in both the state and private sectors, by working to get the incentives right. To be effective it also means empowering children and young people themselves, which in turn requires respecting their rights.

The idea that complex social problems are amenable to cheap technical solutions is the siren song of the software salesman and has lured many a gullible government department on to the rocks. Where ministers buy the idea of a magical software 'solution', as the industry likes to call its products, the outcomes are often disappointing and sometimes disastrous[81]. And the very idea that we can replace police officers, social workers and teachers by ordering Facebook to watch our children and grandchildren more closely is a non-starter. The kids left Facebook years ago for Instagram; they're now headed via Snapchat to TikTok, and to an assortment of gaming platforms.

Finally, universal human rights set the boundaries for state action. Pervasive surveillance, without warrant or suspicion, is contrary to human-rights law, just like torture. Arguments in its favour must be treated with great suspicion and cannot be conceded on utilitarian grounds. Agencies tasked with defending the rules-based international order should defend the basic rights of their own citizens, including the rights of children, rather than seek to undermine them. The rule of law must take precedence over 'national security'. We must maintain a moral advantage over competing authoritarian states, not just a military and technological advantage. End-to-end encryption must therefore remain available for moral reasons. It must also remain for very good cybersecurity reasons – as Levy and Robinson conceded in their earlier paper[82], and as we discussed in "Bugs in our pockets"[83].

**Acknowledgements**

I acknowledge valuable feedback from Sophie van der Zee, John Churcher, Vanessa Teague, Ben Collier, Eileen Munro, Jen Persson, Paul Whitehouse, Susan Landau, Diane Coyle, Richard Clayton, Konstantin Macher, Francis Davey, Alan Cox, Peter Sommer, Javier Ruiz, Nicholas Bohm, Jenny Blessing, Nicholas Boucher, Sam Ainsworth, Mark Seiden, Markus Kuhn, Sam Gilbert, Elina Eickstädt and Ian Levy. Not all of these contributors necessarily agree with all the arguments presented here.

---

[81] Anthony King and Ivor Crewe, 'The Blunders of Our Governments', Oneworld Publications, 2013
[82] Levy & Robinson, "Principles for a More Informed Exceptional Access Debate", Lawfare 2018
[83] Hal Abelson, Ross Anderson, Steven M. Bellovin, Josh Benaloh, Matt Blaze, Jon Callas, Whitfield Diffie, Susan Landau, Peter G. Neumann, Ronald L. Rivest, Jeffrey I. Schiller, Bruce Schneier, Vanessa Teague, Carmela Troncoso, "Bugs in our Pockets: The Risks of Client-Side Scanning", Oct 14, 2021